\documentclass[12pt]{iopart}

\usepackage{iopams} 
\usepackage{graphicx}
\usepackage[ansinew]{inputenc}
\usepackage{array}
\usepackage{color}
\usepackage{latexsym}
\usepackage{cite}

\begin{document}

\title[]{Attosecond Quantum-Beat Spectroscopy in Helium}

\author{Niranjan Shivaram$^{1}$\footnote{Current address: Chemical Sciences Division, Lawrence Berkeley National Laboratory, Berkeley, CA 94720, USA}, Xiao-Min Tong$^2$, Henry Timmers$^1$ and Arvinder Sandhu$^1$}

\address{$^1$Department of Physics, University of Arizona, Tucson, AZ, 85721 USA.}
\address{$^2$Graduate School of Pure and Applied Sciences, and Center for Computational Sciences, University of Tsukuba, Ibaraki 305-8573, Japan.}
\ead{nhshivaram@lbl.gov}

%

\begin{abstract}
The evolution of electron wavepackets determines the course of many physical and chemical phenomena and attosecond spectroscopy aims to measure and control such dynamics in real-time. Here, we investigate radial electron wavepacket motion in Helium by using an XUV attosecond pulse train to prepare a coherent superposition of excited states and a delayed femtosecond IR pulse to ionize them. Quantum beat signals observed in the high resolution photoelectron spectrogram allow us to follow the field-free evolution of the bound electron wavepacket and determine the time-dependent ionization dynamics of the low-lying $2p$ state. 

\end{abstract}

%
%
%
%


Wavepackets form a connection between quantum mechanics and the classical concept of an electron and nuclear motion in atoms and molecules. Due to their importance in physics, chemistry and biology, wavepackets have been studied extensively by both theorists and experimentalists over the past several decades \cite{Zewail1993,Garraway95}. Most studies of electronic wavepackets have focused on high Rydberg wavepackets composed of closely spaced electronic states which have dynamics on 100's of femtoseconds or picosecond timescales \cite{alber1986, parker1986, yeazell1988}. This was mainly because of the lack of short light pulses with enough bandwidth to create and detect fast evolving coherent superpositions of multiple states spanning a wide energy range. With the availability of attosecond pulse trains \cite{Antoine96} and single isolated attosecond pulses \cite{Sansone06} which can have a bandwidth of nearly  a hundred electron volts, it is now possible to create wavepackets composed of excited states \cite{Johnsson07,Choi2010, Mauritsson2010, Shivaram2012prl}, valence states \cite{Goulielmakis2010}, and even core electronic states \cite{Neppl2012} and study their dynamics on attosecond time-scales. 

\begin{figure}
\centering
\includegraphics[width=0.80 \textwidth]{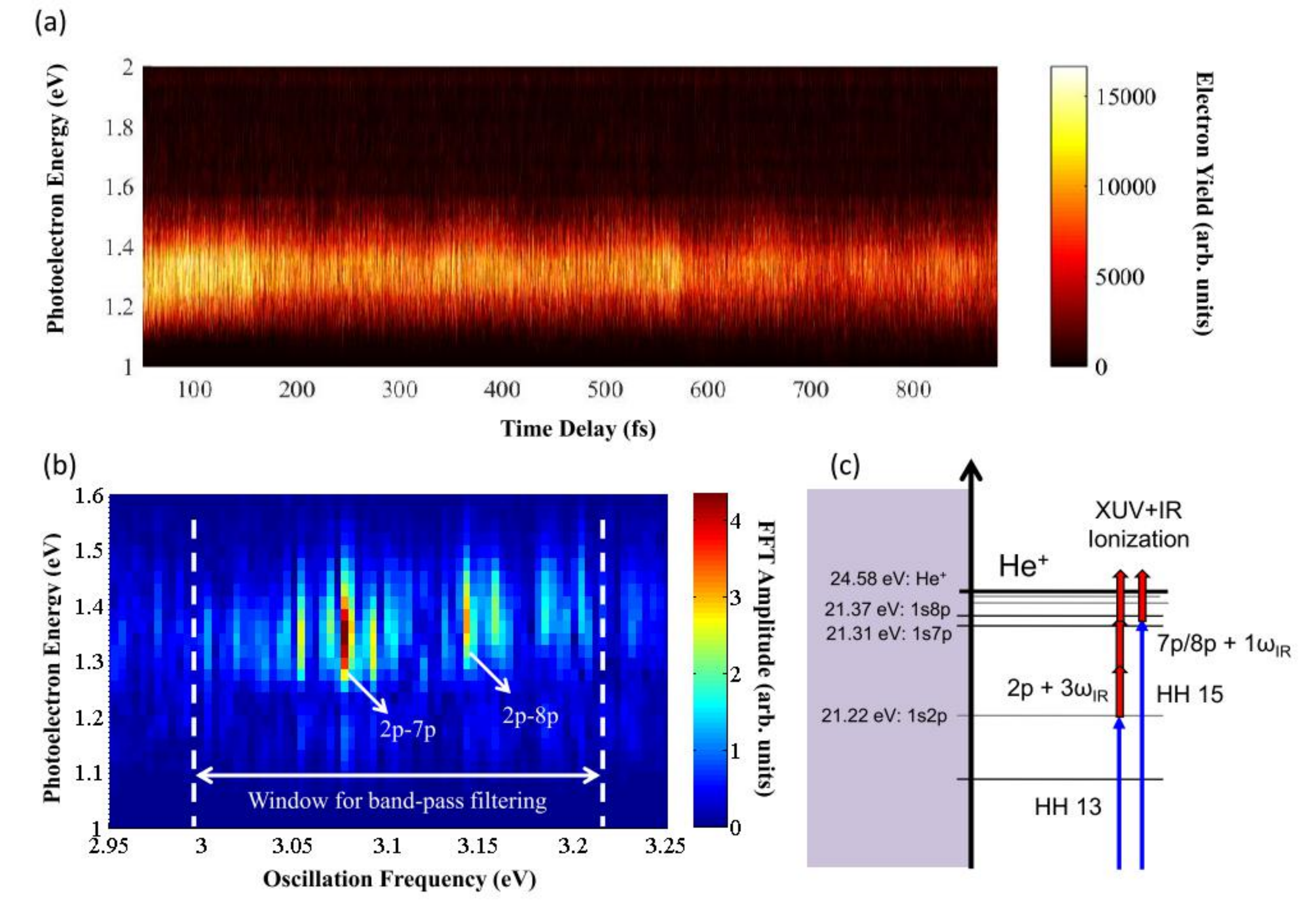}
\caption{(a) Raw experimental photoelectron spectrogram. The IR pulse arrives after the XUV pulse for positive time-delays. (b) Fourier transform of the experimental photo-electron spectrogram. We use units of eV for the oscillation frequency (x-axis) where 3.16 eV corresponds to 2$\omega_{IR}$ or half IR optical cycle (1.3 fs). The main peaks corresponding to interference of photoelectrons arising from the $2p, 7p$ states and $2p, 8p$ states are shown. These dominant interefering paths are shown in (c). The window used for bandpass filtering (see figure \ref{fig:spectrogram_bandpass}) is also shown in (b).}
\label{fig:spectrogram_raw}
\end{figure}

Due to its simplicity and ease of modeling, the Helium atom forms a convenient test bed for the study of electron wavepacket dynamics. Here, we use extreme ultraviolet (XUV) high harmonics in attosecond pulse trains to excite a radial wave-packet of $ 2p, 7p, 8p ..$ states in helium and study its dynamics with attosecond resolution. We use a delayed near-infrared (IR) pulse to photoionize excited states and measure photoelectrons as a function of XUV-IR time delay. We focus on interferences between ionization paths which are related to the quantum beats of the excited electron wavepacket and coupling effects induced by the ionizing IR pulse. Most prior studies have focused on short time evolution \cite{Ranitovic10,Mauritsson2010, Kim2012, Shivaram2012prl, Shivaram2013} or study slow beats over long time scales \cite{Verlet2003, Geiseler2011} except a recent study of anisotropic emission in quantum beat spectroscopy \cite{Lucchini2015}. Here we report unique investigations of field-free dynamics spanning over 100's of femtoseconds with attosecond time resolution and high photoelectron energy resolution. We resolve both the slow wavepacket beating between high Rydberg states and the fast $2p - np$ state beating. From this information we deduce the detailed energy and time dependence of the ionization amplitude of the low-lying 2p state in the wavepacket which has not been explored in previous studies. This approach yields new insight into the complex ionization dynamics of the XUV excited attosecond wavepacket. 


The experimental setup consists of a Ti:Sapphire laser amplifier which produces 45 fs (full-width half max), 2 mJ energy IR pulses at a central wavelength of 785 nm. The laser beam is divided into two arms. In one arm, the laser pulses are focused into a Xe gas filled waveguide to generate an attosecond pulse train consisting predominantly of 13th and 15th harmonics. After using an Aluminum filter to remove residual IR, the high harmonics are focused onto an effusive Helium gas jet where they create excited `$np$' states of the helium atom forming a wave-packet which can be written as (in atomic units)
\begin{equation}
\label{eq:field_free_WP}
\psi(\mathbf{r},t) = \sum \limits_n \psi_{nl}(\mathbf{r})e^{-iE_{nl}t}
\end{equation}
where $l=1$ corresponding to $p$-states and from the energies of the harmonics we know that the sum is over $n=2$ (from HH 13) and $n=7,8$ (from HH 15) as shown in figure \ref{fig:spectrogram_raw} (c). 

\begin{figure}
\centering
\includegraphics[width=0.75 \textwidth]{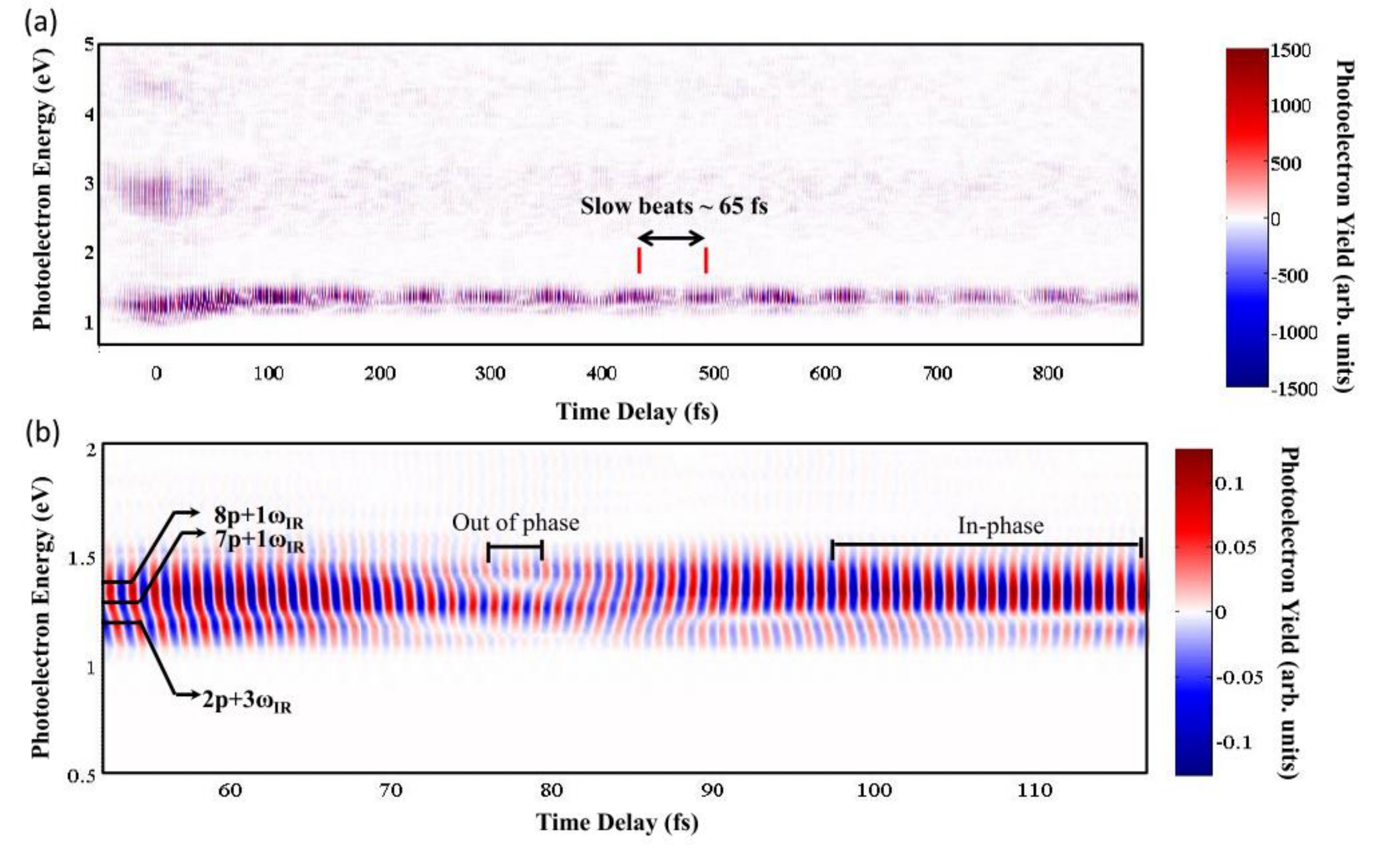}
\caption{(a) Band pass filtered experimental photoelectron spectrogram spanning a time-delay range of -50 fs to 880 fs with 65 attosecond resolution. The photoelectron yield has positive and negative values since we have removed the DC component. The IR pulse arrives after the XUV pulse for positive time-delays. Slow beats with a period of $\sim 65$ fs are clearly observable. (b) A zoom-in of the photoelectron spectrogram in the range 50 fs to 120 fs. Fast oscillations with a period of $\sim 1.3$ fs or $2\omega_{IR}$ frequency are seen. Oscillations at different photo-electron energies have slightly different frequencies leading to in-phase and out-of-phase regions as indicated.}
\label{fig:spectrogram_bandpass}
\end{figure}

In the other arm of the set up, an IR pulse is time delayed and after passing through a lens it is recombined with the HHG beam to propagate collinearly. The XUV and IR pulses have the same polarization. The IR pulse ionizes excited Helium atoms in the focal region at an intensity of $\sim 5 \times 10^{12}$ W cm$^{-2}$ and the photoelectrons produced are imaged using a velocity map imaging setup \cite{Eppink97}. 
The XUV-IR time delay is varied over a large range from -50 fs to 880 fs in 65 attosecond steps where positive delay corresponds to IR arriving after the XUV pulse. The photoelectron data is processed by measuring the electron yield in a narrow window along the XUV and IR polarization direction and this yield is plotted as a function of electron energy and time delay to obtain a photoelectron spectrogram. 

\begin{figure}
\centering
\includegraphics[width=0.75 \textwidth]{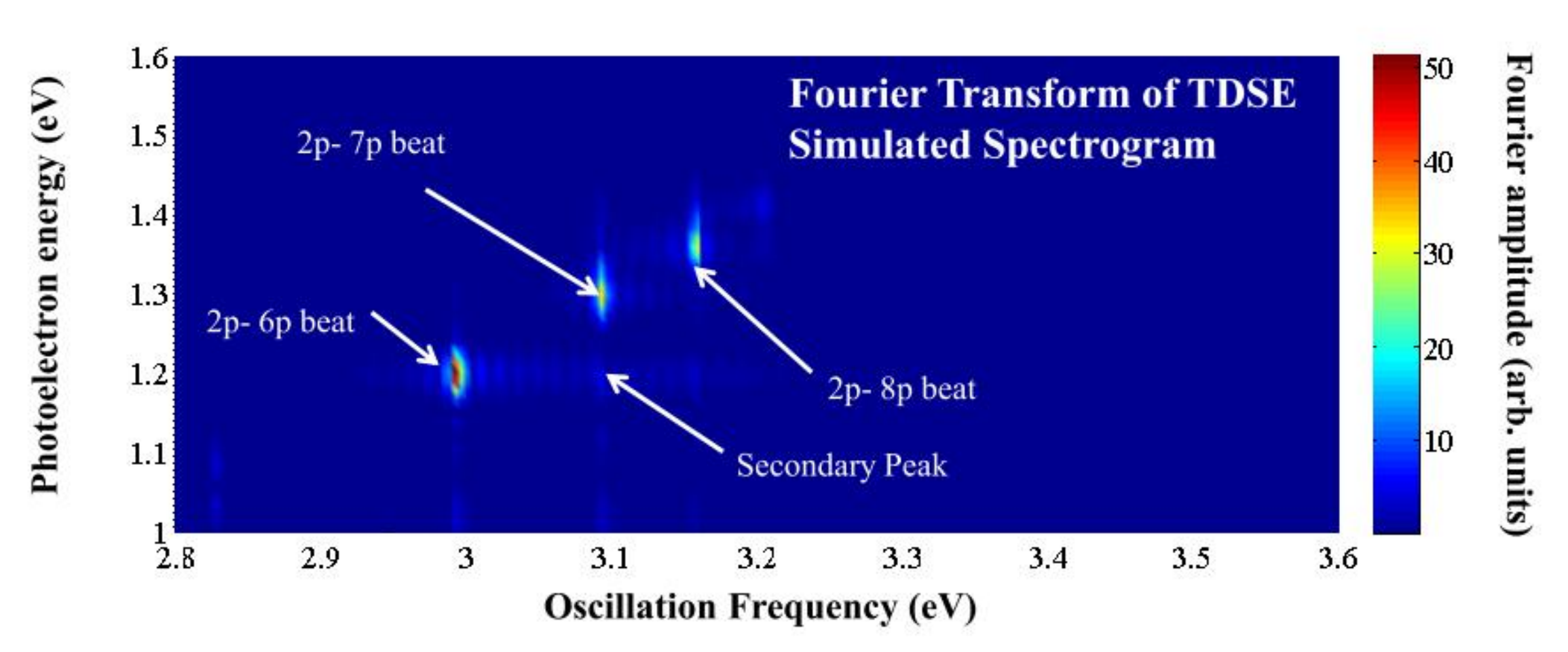}
\caption{The TDSE simulation results showing the Fourier transform of the simulated photoelectron spectrogram. Peaks corresponding to $2p-7p$ and $2p-8p$ interferences are seen.}
\label{fig:spectrogram_sim}
\end{figure}

In the raw photoelectron spectrogram shown in figure \ref{fig:spectrogram_raw} (a), we clearly observe oscillations in the electron yield as a function of time-delay. As we discuss below, the oscillation frequency changes with the electron energy, corresponding to presence of different interference channels. In order to understand the origin of these oscillations we fourier transform the raw data. The Fast-Fourier Transform (FFT) in figure \ref{fig:spectrogram_raw} (b) shows clear peaks corresponding to specific quantum beat signals. We observe two dominant peaks which occur at an oscillation frequency of 3.08 eV and 3.14 eV. An oscillation frequency of 3.16 eV corresponds to $2\omega_{IR}$ frequency or half IR optical cycle period. The peak at 3.08 eV is consistent with a beat occuring due to the interference of photoelectrons from three IR photon ionization of $2p$ state and one IR photon ionization of $7p$ state (Fig.\ref{fig:spectrogram_raw} (c)). The peak at 3.14 eV is consistent with interference of $2p+3\omega_{IR}$  and $8p+1\omega_{IR}$. The locations of these peaks in energy is also consistent with the energy locations of the $2p, 7p$ and $8p$ states. The peaks very close to the $2p-7p$ and the $2p-8p$ peaks are fourier transform artifacts, though the peak near 3.18 eV may arise from $2p-9p$ interference. Since $2p-7p$ and $2p-8p$ interferences have a dominant contribution, we also expect a slow quantum beat of $\sim 65$ fs corresponding to the energy gap between $7p$ and $8p$ states.

In order to focus on specific channels with greater detail, we remove the noise and the DC value of the electron yield by band-pass filtering with a window around prominent peaks in the FFT shown in figure \ref{fig:spectrogram_raw} (b). The photo-electron spectrogram resulting from the fourier bandpass filtering is shown in figure \ref{fig:spectrogram_bandpass} (a). Since the data in figure \ref{fig:spectrogram_bandpass} (a) is for a time-delay scan of $\sim 1$ ps, the features and variations on attosecond timescales are not visible. However, we see clear beats which occur with a period of $\sim 65$ fs that agrees well with the energy difference between $7p$ and $8p$ states discussed above. Though this plot shows a scan of a time-delay range that includes the temporal overlap region between the XUV and IR pulses, in the following we focus only on the dynamics in the non-overlap region since we are interested in field free wavepacket dynamics.

Figure \ref{fig:spectrogram_bandpass}(b) shows the same experimental data as in figure \ref{fig:spectrogram_bandpass}(a) but zoomed in to show fine features in the range of 50 fs to 120 fs. First, we see fast oscillations in the electron yield as a function of time delay and these oscillations occur with a period of $\sim 1.3$ fs which is half an IR optical cycle ($2\omega_{IR}$ frequency). More interestingly, we see that the oscillatory behavior is not the same at all photoelectron energies. There appears to be a slight variation in oscillation frequency at different energies between 1.1 eV and 1.5 eV. This results in these oscillations going out-of-phase and coming into phase as indicated in figure \ref{fig:spectrogram_bandpass}(b). Also shown in figure \ref{fig:spectrogram_bandpass}(b) are the expected central energies of photoelectrons arising from IR induced ionization of unperturbed XUV excited $2p, 7p$ and $8p$ states.

We also performed time-dependent Schrodinger equation (TDSE) simulations  to model the wavepacket dynamics observed in our experiments with a single-active-electron model potential \cite{Tong05c}. Using XUV and IR parameters similar to the experiments, we numerically obtained photoelectron spectrograms \cite{Tong10,Tong10b} that also exhibit fast $2\omega_{IR}$ oscillations along with dephasing and rephasing of beat signals in good agreement with the experimental results. In figure \ref{fig:spectrogram_sim} we plot a Fourier transform of the simulated spectrogram. The peaks corresponding to the quantum beats between $2p-6p, 2p-7p$ and $2p-8p$ stand out distinctly. The comparison with experimental figure \ref{fig:spectrogram_raw}(b) is good except that the peak due to $2p-6p$ interference is indistinguishable in the experiment, probably due to small differences between simulation and experimental parameters stemming from the uncertainities in exact widths and amplitudes of XUV harmonics. In the simulation (Fig. \ref{fig:spectrogram_sim}), we observe a secondary peak around 1.2 eV photoelectron energy below the $2p-7p$ and $2p-8p$ peaks, with a minima in signal levels around 1.25 eV photoelectron energy. This secondary peak feature is also seen in the experimental data (Fig. \ref{fig:spectrogram_raw} (b)) at 1.15 eV photoelectron energy. 

\begin{figure}
\centering
\includegraphics[width=0.85 \textwidth]{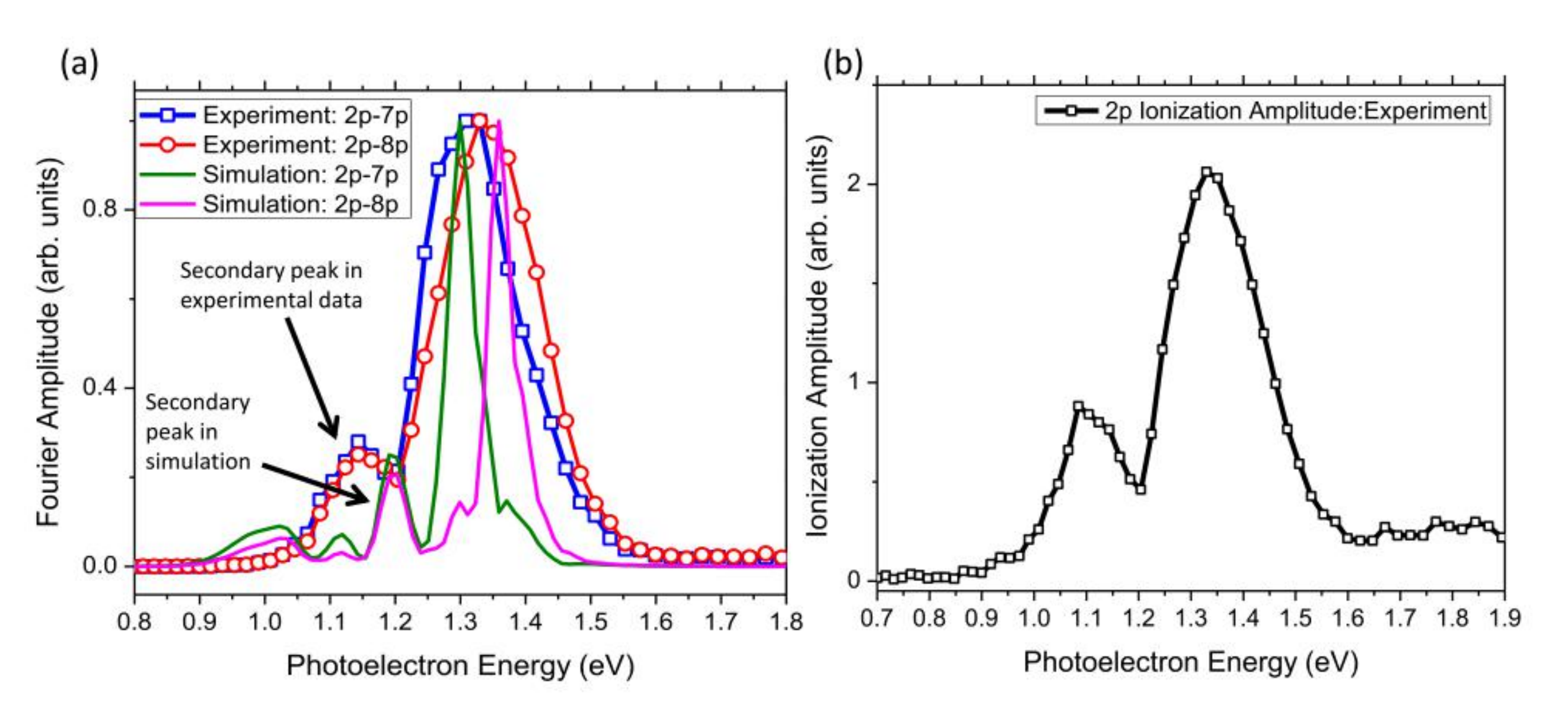}
\caption{ (a) Line-outs of the fourier transform of experimental and simulated photoelectron spectrograms taken at the oscillation frequency corresponding to $2p-7p$ and $2p-8p$ beats. A minimum and a secondary peak are seen in the $2p-7p$ and $2p-8p$ line-outs both in the experimental and simulated data. The energy location of these features in experiment differ from those in simulation by $\sim 50$ meV though the main peaks agree well. (b) The $2p$ ionization amplitude extracted from the experimental data (see text for details).}
\label{fig:spectrogram_lineout}
\end{figure}

In order to quantitatively compare the quantum beat signals in the experiment and simulation, in figure \ref{fig:spectrogram_lineout}(a) we show photoelectron energy line-outs of figure \ref{fig:spectrogram_sim} taken at the frequency of the $2p-7p$ and $2p-8p$ beats along with the corresponding line-outs of the experimental data from figure \ref{fig:spectrogram_raw}(b). The locations of the main peaks in experimental and simulation lineouts differ by less than $\sim 20$ meV. The energy dependence is also very similar; as both experiment and simulation data exhibit a dip and secondary peak at lower photoelectron energy, but the location of secondary peaks shows a slight mismatch ($\sim 50$ meV). This energy dependent beat signal can be used to extract the amplitudes of the ionization matrix elements for various excited states composing the wavepacket.

\begin{figure}
\centering
\includegraphics[width=0.85 \textwidth]{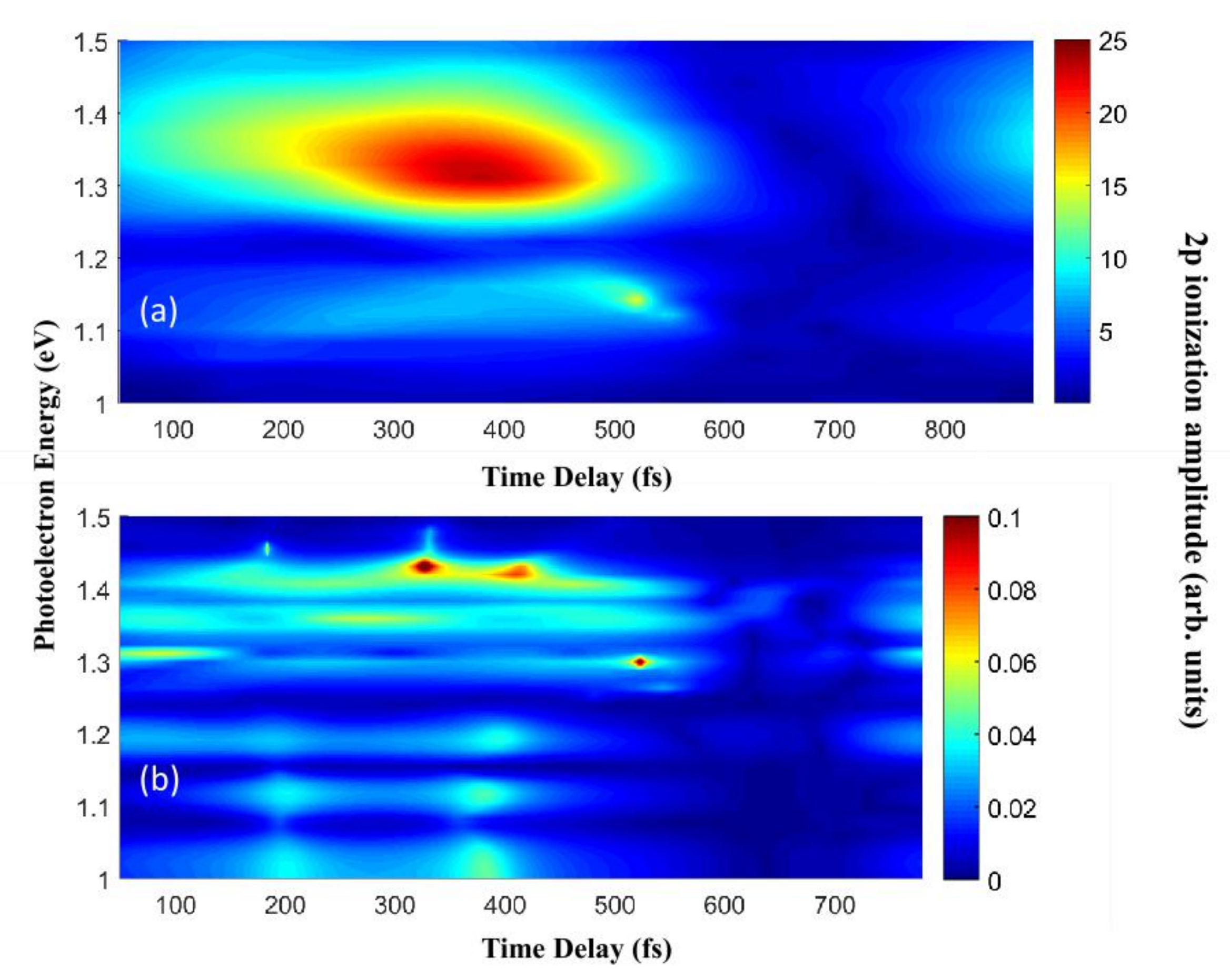}
\caption{Ionization amplitude of $2p$ as a function of photoelectron energy and time-delay between XUV and IR pulses. The ionization amplitude is obtained from the photoelectron spectrogram by applying a fourier transform, followed by an inverse fourier transform as discussed in the text. (a) Experimental data (b) Simulation.}
\label{fig:Ion_prob_2p_vs_time_delay_3D}
\end{figure}

The IR pulse induced ionization of the $2p, 7p, 8p$ states leads to interferences in the continuum and various oscillatory components of the time and energy dependent electron yield can be written as  
\begin{equation}
\label{eq:IonAmp}
I_{mn}(\epsilon,\tau) = A_{mp}(\epsilon, \tau)A_{np}(\epsilon, \tau)cos(\omega_{mp,np}\tau+\phi_{mp,np})
\end{equation}
where $A_{mp}(\epsilon, \tau)$ and $A_{np}(\epsilon, \tau)$ are amplitudes of the ionization matrix element (henceforth referred to as ionization amplitudes) from the interfering $mp$ and $np$ states, which depend on photoelectron energy $\epsilon$ and slowly vary with time delay $\tau$. $\omega_{mp,np}$ represents the energy difference between the $mp$ and $np$ states (in atomic units), $\tau$ is the time delay between XUV and IR pulses and $\phi_{mp,np}$ is the phase difference between the ionization matrix elements from the $mp$ and $np$ states. The lineouts shown in figure \ref{fig:spectrogram_lineout}(a) are thus a measure of the amplitude of two dominant $I_{mn}(\epsilon,\tau)$ terms corresponding to products $A_{2p}A_{7p}$ and $A_{2p}A_{8p}$. 
When this information is combined with the amplitude of slow beating between 7p and 8p which gives us $A_{7p}A_{8p}$, we can extract the individual ionization amplitude for the low-lying $2p$ state which is not accessible through other methods. The slow time delay dependence of the $2p$ ionization amplitude will be discussed later and is ignored in this analysis.
The result of this excercise is shown in figure \ref{fig:spectrogram_lineout}(b). The ionization amplitude of $2p$ also exhibits a double peak structure similar to the Fourier amplitudes, albeit the amplitude of the secondary lower energy peak is now quite significant compared to the main peak. We discuss the physical origin of the secondary peak below.

During multiphoton above-threshold ionization (ATI) of atoms, new structures and enhancements have been observed in photoelectron peaks \cite{Freeman1987, Nandor1999,Potvliege2009}. These structures have been attributed to field-dressed/ponderomotively shifted states which are resonant with the ground state on the way to the continuum during ionization and are known as Freeman resonances \cite{Freeman1987}. We believe that the secondary lower energy peak in the ionization amplitude of $2p$ observed here is due to field-dressed states that are 2-photon resonant with the $2p$ state. From selection rules these states are of p and f character. We identify that the $nf$ and $np$ states have an almost equal contribution to the secondary peak by removing the $np$ and $nf$ states near $7p$ and $8p$ in the TDSE simulations. In addition, we find that Freeman resonances corresponding to $n=6$ states play a dominant role in creation of this structure in $2p$ ionization amplitude. 

\begin{figure}
\centering
\includegraphics[width=0.75 \textwidth]{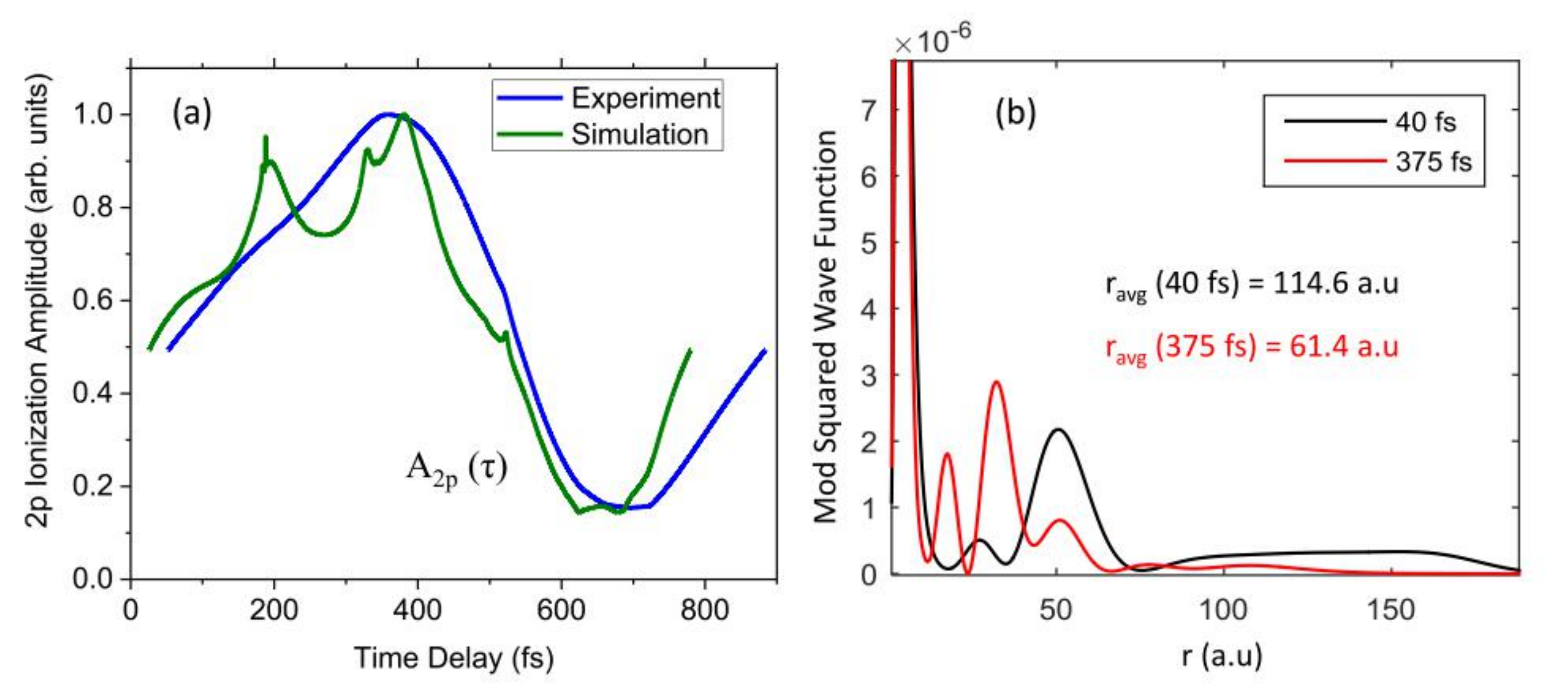}
\caption{(a) Data from figure \ref{fig:Ion_prob_2p_vs_time_delay_3D} integrated over photoelectron energies. The experimental and simulated ionization amplitudes for $2p$ agree very well. (b) Simulation results for the modulus squared radial wavefunction of the XUV excited wavepacket at two different time delays corresponding to low ionization amplitude (40 fs) and high ionization amplitude (375 fs). The average distances ($r_{avg}$) of the radial wavepacket are also shown.}
\label{fig:Ion_prob_2p_2D}
\end{figure}

We now focus our attention on studying the time-dependent dynamics of the XUV excited wave-packet. In previous studies \cite{alber1986, parker1986, yeazell1988}, it has been shown that when high lying Rydberg wave packets are excited, they periodically move close to the core and back. Ionization only occurs when the wavepacket is close to the core because the electron-parent core interaction is larger. Here, we are interested in the ionization dynamics of the low lying $2p$ state in the XUV excited wave packet. We extend the analysis done above where we obtained the energy dependent ionization amplitude of the $2p$ state to obtain the ionization amplitude of $2p$ as a function of time. This is done by performing a Fourier transform on the photoelectron spectrogram, gating the mp-np beat peak followed by an inverse fourier transform to obtain the energy and delay dependent products $A_{2p}(\epsilon, \tau)A_{7p}(\epsilon, \tau)$, $A_{2p}(\epsilon, \tau)A_{8p}(\epsilon, \tau)$ and $A_{7p}(\epsilon, \tau)A_{8p}(\epsilon, \tau)$ from which we obtain $A_{2p}(\epsilon, \tau)$. Figure \ref{fig:Ion_prob_2p_vs_time_delay_3D} (a) shows the experimental $2p$ ionization amplitude as a function of photoelectron energy and time-delay. We apply the same procedure to the simulated photoelectron spectrogram and the $2p$ ionization amplitude obtained is shown in figure \ref{fig:Ion_prob_2p_vs_time_delay_3D} (b). 

The simulation plot figure \ref{fig:Ion_prob_2p_vs_time_delay_3D} (b) has a lot more structure than the experimental plot in figure \ref{fig:Ion_prob_2p_vs_time_delay_3D}(a). This difference is mainly due to the lower energy resolution and noise in the experiment. The overall temporal behavior of experimental ionization amplitude agrees very well with the TDSE calculation. To see this better, we integrate the data in figure \ref{fig:Ion_prob_2p_vs_time_delay_3D} along the photoelectron energy axis and plot it in figure \ref{fig:Ion_prob_2p_2D} (a). The ionization amplitude of the $2p$ state varies as a function of time delay by nearly a factor of 5 between maximum and minimum points which are $\sim$ 325 fs apart. In figure \ref{fig:Ion_prob_2p_2D} (b), we show the TDSE calculations of the modulus squared radial wavefunction of the XUV excited wavepacket corresponding to delay times of 40 fs and 375 fs. The average radial distances of the wavepacket are also shown. It is apparent from these plots that the wave packet moves closer to the core at 375 fs, which in turn leads to higher ionization amplitude observed in experiments. This long time scale in the wavepacket motion most likely arises from the presence of other Rydberg states in the excited wavepacket which do not interfere with the 2p ionization channel in the continuum. It is very interesting that even a low lying state like $2p$ shows such a large sensitivity to the radial motion of the wavepacket during ionization. We have thus followed the evolution of a radial wavepacket and measured the effect of radial motion on the ionization probability of the low lying $2p$ state in the wavepacket.

In conclusion, we have performed a high resolution attosecond quantum beat spectroscopy measurement in helium using XUV and IR pulses. The XUV excited wavepacket predominantly consisting of $2p,7p$ and $8p$ states was probed by a time delayed IR pulse on a time scale close to a picosecond with attosecond resolution. The photoelectron yield was measured as a function of energy and XUV-IR time delay and beats between the $2p, 7p$ and $8p$ states due to interferences between multiple paths to the continuum are clearly observed. Using fourier analysis, we observed structures in the photoelectron energy dependent ionization amplitude of the $2p$ state and attributed them to Freeman resonances. Finally, we used a time-frequency analysis technique to extract the time dependent ionization amplitude of the low lying $2p$ state which shows a high sensitivity to the radial motion of the wavepacket. Extending these measurements to study electronic wavepackets in molecular systems can yield detailed information about the dynamics of such wavepackets, especially on the interplay between electronic and nuclear degrees of freedom \cite{Ranitovic14}.






\section*{Acknowledgements}
This work was supported by the National Science Foundation (NSF) under contract PHY-1505556. XMT was supported by a Grant-in-Aid for Scientific Research (C24540421) from the Japan Society for the Promotion of Science and HA-PACS (Highly Accelerated Parallel Advanced system for Computational Sciences) Project for advanced interdisciplinary computational sciences by exascale computing technology.

\section*{References}
\bibliographystyle{iopart-num}
\bibliography{RadialWP_Refs}

\end{document}